\documentclass{ws-spin}
\usepackage{multicol}
\begin{document}

\catchline{11}{4}{2021}{}{}
\markboth{A. Useinov}{Application of the point-like contact model..}

\title{Application of the point-like contact model: \\Resistance Oscillations of the Domain Wall in Magnetic Nanowires and Junctions due to Mean Free Path Effects}

\author{Artur Useinov}

\address{International College of Semiconductor Technology, National Yang Ming Chiao Tung University\\ 1001 Ta Hsueh Rd, Hsinchu, 30010, Taiwan R.O.C.\\
\email{artu@nycu.edu.tw}
}
%

\maketitle

\begin{history}
\received{20 October 2021}
\accepted{14 December 2021}
\end{history}

\begin{abstract}
This work is focused on determining the electrical resistance, which induced by single domain wall in magnetic nanowire with a negligible defect. The provided model covers a wide range of nanowire’s diameters. The obtained result demonstrates a few orders rapid reduction of the domain wall resistance accompanied by its possible deviations {\it{versus}} the diameter growth ranging from 1.2~nm to 15.2~nm. The origin of these deviations, which are also identified as oscillations, is referred to the non-uniform electron scattering on the domain wall due to the intermixing electron scattering conditions: ballistic for one spin channel and quasi-ballistic for other one with opposite spin direction. It may happen when the domain wall width by value is approximately in between two lengths: a mean free path with the spin down and spin up. The indirect evidence of this finding is also coming from the fact that homogeneous nanowires shows the most valuable domain wall resistance oscillations by magnitude rather than segmented magnetic nanowires. In addition to the approach, where DW width is constant, the other reasonable model is used when the domain wall can be constrained for some conditions. The same results are valid for magnetic junctions with domain wall. Finally, resistance simulation in the diffusive range, when a diameter of the nanowire (or point-like junction) is larger than any of spin resolved mean free path of electrons, successfully follows experimental data for the single and double domain wall resistances available in literature.
\end{abstract}

\keywords{Nanomagnetics, magnetic domain wall, resistance, ballistic magnetoresistance, nanowires.}

\begin{multicols}{2}

\section{Introduction}

The control of magnetic domain wall (DW) motion and resistive states are key points for electronic applications. For example DW can be utilized for a computational logic\cite{allwood_magnetic_2005}, energy-effective memristor\cite{lequeux_magnetic_2016}, magnetic synapse for neural network\cite{kaushik_comparing_2020}, working cell of the magnetic DW racetrack memory\cite{parkin_magnetic_2008}, {\it{etc}}. Natural lattice defects, artificially created interfaces between different segments in magnetic nanowires (NWs)\cite{mohammed_angular_2015,mohammed2017} as well as geometrical modulations\cite{chandra_sekhar_depinning_2014} are potential centers for a DW trapping. Numerous theoretical and experimental works focused on the investigation of the dynamical motion of the single DW\cite{piao_intrinsic_2013,yan_beating_2010} and its resistance impact\cite{levy_resistivity_1997,tatara_resistivity_1997,tatara_theory_2000,hassel_resistance_2006}. One of these theoretical works generated a model of the total resistivity of the thin ferromagnetic films showing that both transversely and longitudinally oriented DWs are additional source of the resistivity\cite{levy_resistivity_1997}. In contrast, focusing on the ferromagnetic wires with DWs, another work showed that resistivity can be decreased by the reason of electron decoherence due to DW nucleation\cite{tatara_resistivity_1997}.  In addition, application of the semiclassical model\cite{useinov_giant_2007,useinov_spin-resolved_2020}  of the point contact (PC) allows to estimate giant ballistic magnetoresistance (${\rm{BMR}}$)\cite{garcia_magnetoresistance_1999} as well as single DW resistance in magnetic nanojunctions and nanowires, where the resistance due to electron scattering on the DW only increases.

The novel specific DW state, which is almost independent of the material parameters and located in a ferromagnet within a nanoconstriction, was discovered by Bruno\cite{bruno_geometrically_1999}. The width of this DW type is conjugated with the dimension of the constriction and can be much smaller in contrast to the well-known Bloch- and N\'eel- DW types. Bruno made the numerical estimations of the DW width and energy, showing the evidence of the giant BMR in nanoscale PCs. Furthermore, after Bruno's predictions and further model development by Yan {\it{et al.}}\cite{yan_beating_2010}, experimental detection of the novel Bloch-point DW was obtained by Da Col {\it{et al.}}\cite{da_col_observation_2014} in a cylindrical magnetic NWs. It was claimed that Bloch-point DW is the zero-mass nanomagnetic object with topologically protected state and transverse magnetic geometry, having a unique dynamic properties. The transverse ordering of the magnetizations (along NW) is more preferred magnetic state in NWs with diameter $d_0<$250~nm rather than a vortex state, observed in the work\cite{wong_current-induced_2016} for the case $d_0>$250~nm, where magnetization are curling around the wire axis. The prevention of the vortex states is kept due to domination of the exchange energy against dipolar and anisotropy energies\cite{da_col_observation_2014,wieser_domain_2004}. 
When the lateral dimension of the NW achieves 40~nm, or less, the physics of the electron transport is changed due to quantum confinement. The goal of present work is calcuation of the DW resistance and its behavior at the confined geometry, assuming that magnetizations ($M$) are oriented along magnetic NW  and DW is characterized as linear slope of $M$ projections. The dimensional factor becomes dominated, since mean free paths (MFP) of the spin-resolved carriers are now comparable with DW width. For simplicity, the possible NW anisotropy change is omited from consideration. 

\begin{figure*}[!t]
\centering
\includegraphics[clip=true, width=6.0in]{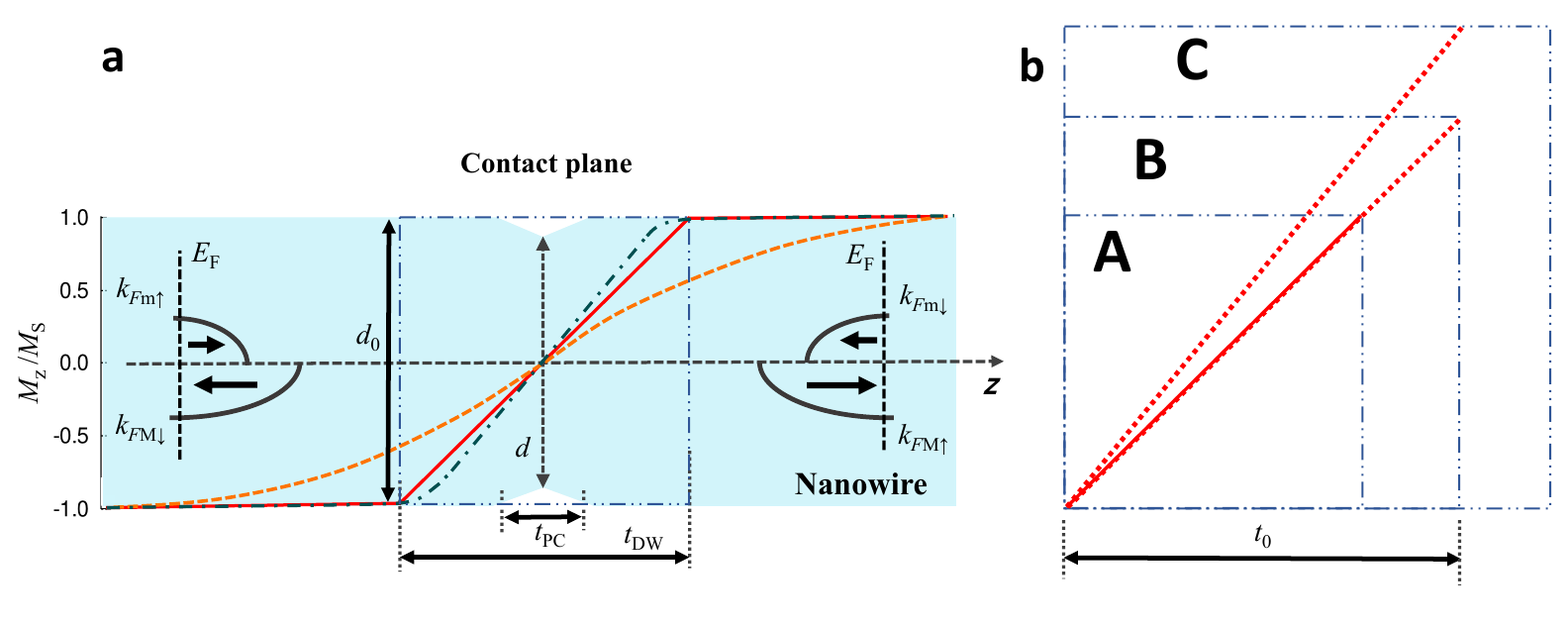}
\caption{Magnetization projections of DWs in nanowire: \textbf{a}  The geometrically constrained DW with magnetic tail as the most relevant model of the constrained state is shown as green dash-dot curve; The constrained ($t_{\rm{DW}}=d$) tail-removed linear projection as a simplified DW approach of the present model is shown as red solid line; And alternative magnetic DW profile is shown as orange dashed line. The left and right sketches show the majority and minority electron subbands, respectively. $M_{\rm{S}}$ is saturation magnetization, $M_{\rm{Z}}$ - projection of the spatial magnetization vectors on {\it{z}} axis, where {\it{z}} is a quantization axis along the electron transport direction. \textbf{b} The tilting of the $M_{\rm{Z}}/M_{\rm{S}}$ depending on the longitudinal cross section of the NW in form of the square with diameter $d_0$, where A and B cases correspond to the constrained DW approach (here tilting is the same), C is the unconstrained case, $t_{\rm{DW}}= t_0$, where the tilting is changed due to increased $d_0$}
\label{fig1}
\end{figure*}

\section{Theoretical Model}

The main benefit of the present model is the adaptation of the PC model\cite{useinov_spin-resolved_2020} for DW resistance ($\Delta R$) simulation in magnetic NWs within an approach of the tail-removed linear DW approach, Fig.\ref{fig1}a. Whatever the DW's 3D magnetization direction behavior is, the most important for an electron transport is its projection on the axis of the transport direction {\it{z}}. For simplicity, the red solid line is adjusted as such projection for the present model and, at the same time, the red curve is a linear potential energy profile which spin-down electron has along {\it{z}}. Spin-up electron has a mirror reflected potential in relation to {\it{z}} axis due to exchange energy, according the Stoner model. To explain $\Delta R$ behavior in NW, it is assumed, that NW contains a small idealized defect, which is considered as PC, and diameters of NW and PC are almost the same values: $d_0 \approx d$, that make DW-induced potential energy profile a valuable contributor to the PC's resistance in addition to other source - the resistance related with PC's geometrical constriction due to MFP effects \cite{useinov_mean-free_2007}. The same model will work for $\Delta R$ coming from DW located nearby PC in magnetic junction without an existence of NW (at this case, there is only $d$, and $d_0$ is out of consideration). 

The possible most exact dependences of the DW projections are presented in Fig.\ref{fig1}a: $M_{\rm{Z}} \left( z \right) = M_{\rm{S}}  \times \tanh \left( {z/t_{{\rm{DW}}} } \right)$ and $M_{\rm{Z}} \left( z \right) = M_{\rm{S}} (2/\pi )\arcsin \left[ {\tanh \left( {2 z/t_{{\rm{DW}}} } \right)} \right]$ as green dash-dot and orange dashed curves, respectively. They are not considered here for simulations due to complicated solution of the Schr\"odinger equation. In present model both these profiles are exchanged on the linear slope profile $M_{\rm{Z}} \left( z \right) = M_{\rm{S}} \times z/{ t_{{\rm{DW}}} }$ shown by red solid curve. In terms of electron potential energy it corresponds to $U(z) \approx E_{{\rm{ex}}} z/t_{{\rm{DW}}}$, where $E_{{\rm{ex}}}$ is exchange energy in ferromagnet. Furthermore, the condition of the Bruno’s model \cite{bruno_geometrically_1999}, where the DW width ($t_{\rm{DW}}$) strongly correlates with $d$ at the restricted geometry $t_{\rm{DW}} \approx d$ is important. The DW width becomes constant when $d_{0}$ increases and overpass some dimensional threshold $t_{0}$, Fig.\ref{fig1}b, and the tilting of the projection profile increases at this case.

\begin{figure*}[!t]
\centering
\includegraphics[clip=true, width=6.8in]{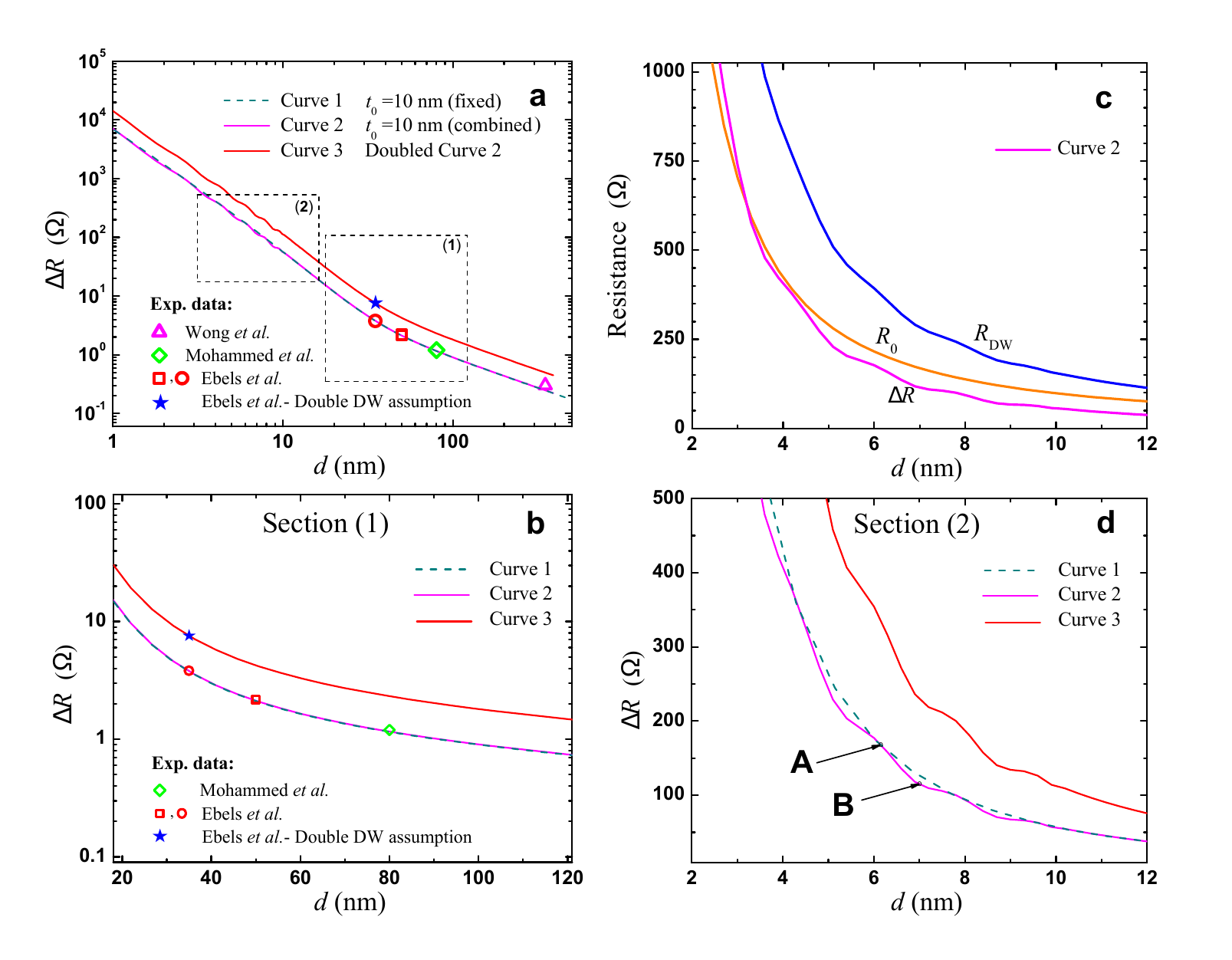}
\caption{Domain wall resistance $\Delta R$ {\it{vs.}} diameter of the PC, $d \to d_0$. \textbf{a} Curves 1 and 2 correspond to the cases of the $\Delta R$ at fixed $t_{{\rm{DW}}}=t_0=10$~nm, and the one for the combined case: $t_{{\rm{DW}}} = d$, while $t_{{\rm{DW}}} = t_0$, at $d > t_0$, highlighting $\Delta R$ oscillations with $d$ and how the curves are matched with experimental data. Curves 3 shows the doubled DW resistance $2 \Delta R$ with $d$. Other parameters are common for these curves: $\ell_\uparrow ^{L(R)} = 12$~nm, $\ell_ \downarrow ^{L(R)}  = 3.0$~nm; $k_ \uparrow^L  = 0.61$~\AA$^{-1}$, $k_ \downarrow ^L  = 1.08$~\AA$^{-1}$, $k_ \downarrow ^R  = 1.0801$~\AA$^{-1}$ and $k_ \uparrow ^R  = 0.6101$~\AA$^{-1}$. Panel \textbf{c} shows $R_{{\rm{DW}}}$ and $R_{0}$ and resulting $\Delta R=R_{{\rm{DW}}}-R_{0}$ at linear scale, where the oscillations are most visible. Panels \textbf{b} and \textbf{d} show a zoom of sections (1) and (2), respectively. \textbf{d} Two points A and B demonstrate the relative 10\% deviation of the curves 1 and 2 from each other. }
\label{fig2}
\end{figure*}

In terms of applications, the problem is actual for a system with large amount of magnetic NWs (PCs) having distribution by $d_0$ ($d$), {\it{i.e.}} the set of the NWs (PCs) with varies $d_0$ ($d$), or for a single cone-like NW, which has a periodicaly located defects. The present model is applicable for homogeneous as well as for segmented magnetic NWs, and the DW is pinned in the center of the PC, or on the interface between segments, respectivly. In real cases, DW center can stuck nearby defect location (out of the PC's center), but still the model will be valid due to the proximity of the systems. The lateral dimension $t_{\rm{PC}}$ of the PC, along {\it{z}} axis, is assumed to be vanishing. The contact plane is keeping validity of the quantum boundary conditions for an electron transport\cite{useinov_spin-resolved_2020,useinov_mathematical_2020}. The segment’s resistance of the NW can be estimated roughly {\it{via}} classical equation $R_{{\rm{Seg}}}  = 4\rho L_{{\rm{Seg}}} /(\pi d_0^2 )$, where $\rho$ is the resistivity of the material at low dimension, $L_{{\rm{Seg}}}$ is a segment length. The total resistance of the system is $R_{{\rm{tot}}}  = R_{{\rm{Seg 1}}}  + R_{{\rm{Seg 2}}}  + R_{{\rm{PC}}}$. The considered $\Delta R$ is the difference between $R_{{\rm{tot}}}$ with DW, where $R_{{\rm{PC}}}  \equiv R_{{\rm{DW}}}$, and $R_{{\rm{tot}}}$ without one, where $R_{{\rm{PC}}}  \equiv R_{\rm{0}}$, {\it{i.e.}} $\Delta R  = R_{{\rm{DW}}}  - R_0$, where resistances from segments cancel each other. As a result, it is not necessary to consider the resistance of NW itself, the main focus of the present study is $R_{{\rm{PC}}}  = V/I_{{\rm{PC}}}$ which have to be derived with and without DW. According to Useinov-Tagirov model\cite{useinov_spin-resolved_2020}, the total charge current $I_{{\rm{PC}}}  = I_ \uparrow   + I_ \downarrow$ is:
\begin{equation}
\label{eq1}
I_{{\rm{PC}}}  = \frac{{2e^2 }}{h}\frac{{AV}}{{2\pi }}\sum\limits_{s  =  \uparrow , \downarrow } {k_{{\rm{F}}\;{\rm{min,}}s }^2 \int_0^\infty  {dk} \frac{{J_1^2 \left( {ka} \right)}}{k}} F_s  (k)
\end{equation}
where $J_1 \left( k a \right)$ is Bessel function; the voltage-dependent transport term {\small $F_s  (k) = \left\langle {\cos \left( {\theta _{L,s} } \right)D_s  } \right\rangle - {N_{1,s}(k) \left\langle {\cos \left( {\theta _{L,s} } \right)W_{L,s} (k)} \right\rangle  - N_{2,s}(k) \left\langle {\cos \left( {\theta _{L,s} } \right)W_{R,s}(k) } \right\rangle } $} contains a ballistic ($d \ll \ell _s $) and diffusive-responsible ($d \gg \ell _s$) terms; The corner brackets are averaging over incident angle of the electron trajectory $\theta _{L,s}$. In present model the possible quantization of the $\theta _{L,s}$ at the small dimensions is neglected for simplicity. Last two terms are responsible for diffusive and quasi-ballistic transport, they are sensitive to the spin-resolved MFP of electrons $ \ell _s$, they include integrals $N_{1(2)}$ and $W_{L(R)}$, which contain quantum mechanical transmission coefficient (QTC) of the system $D_s  = D_s \left( {\theta _{L,s} ,V, k^{L}_{s}, k^{R}_{s}, d} \right)$ and radial variable $k$; Other parameters: $A = {\textstyle{1 \over 4}}\pi d^2$ is cross-section area of the PC, having a radius  $a = {\textstyle{1 \over 2}} d$ , $k_{{\rm{F}}\,\min ,s}$ is minimal spin-dependent Fermi wavenumber among $k_s^L $ and $k_s^R$ of the left (right) sides of the PC, $s$ is a spin index. The assumption $D_s \left( {\theta _{L,s} ,V,k^{L}_{s}, k^{R}_{s},d} \right) \to D_s \left( {\theta _{L,s}, V \to 0,k^{L}_{s}, k^{R}_{s},d} \right)$ is taken as reasonable for the DW simulation at the small positive voltage $V$, left side of NW (PC) is grounded. According to Bruno\cite{bruno_geometrically_1999}, the unconstrained DW width threshold is $t_0  = 2\sqrt {A_{ex} /K_{un}}$, where $A_{ex} $ and $K_{un}$ are exchange stiffness constant and uniaxial magnetocrystalline anisotropy constant in metal. For example, it gives $t_0  = 10.6$~nm at $A_{ex}=1.4\times10^{-6}$~erg/cm and $K_{un}=5.0 \times10^6$~erg/cm$^3$ for Co NW, borrowing material parameters from Ebels {\it{et al.}} \cite{ebels_spin_2000}. Applying these estimations, the condition of the constrained DW is used as a main approach, here $t_{\rm{DW}}=d$ at $t_{\rm{DW}} < t_0$. Another words, $t_{\rm{DW}}$ is a varying parameter of the system, which approximately grows up with $d$ until $t_0$ is achieved. The condition when $d$ is comparable with one of the spin-dependent MFP is important for the considered  $\Delta R$ behavior. The wide range of $t_0$ is theoretically considered here: $t_0 = 10$~nm, $t_0 = 16$~nm and $t_0 =36$~nm. Since the spin diffusion length is around 60~nm in Co, that is large enough than $t_0$, an impact of the electron spin flip leakages is out of present consideration. The exact analytical solution for the spin-dependent QTC, which is base on Airy functions for the sloped potential and available in Ref.\cite{useinov_giant_2007}, was used for the tail-removed DW, characterizing the electron scattering from ballistic to diffusive conditions. The considered QTC is applicable for two spin channels with $D_ \uparrow$ and $D_ \downarrow$, while the case without DW corresponds to the condition when $D_{ \uparrow ( \downarrow )}  = 1.0$. The approach of small $V$ in the model of the symmetric (homogeneous) NW assumes that $k_s^R$ are taken as voltage independent, but slightly larger than $k_s^L$ for the negligibly small positive $V$.
\begin{figure*}[!t]
\centering
\includegraphics[clip=true, width=4.8in]{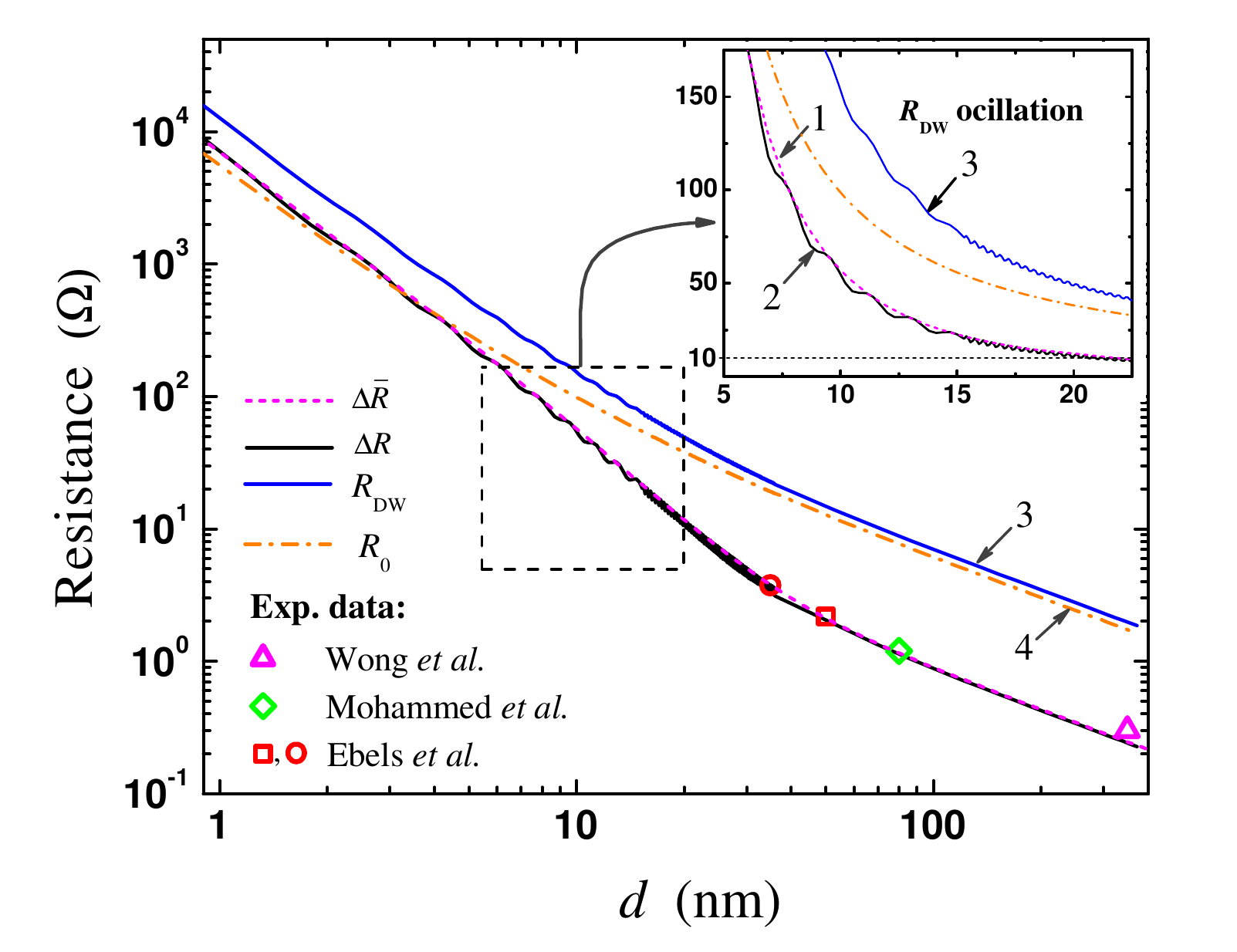}
\caption{Single DW resistance {\it{vs.}} diameter of the PC, that approximately equals to the DW resistance in NW at the condition $d \to d_0$. Curves 1 correspond to the case of the $\Delta \bar R$ at $t_{{\rm{DW}}}=t_0=10$~nm, and curve 2, $\Delta R=R_{{\rm{DW}}}-R_0$, is derived for the combined case of the DW width: $t_{{\rm{DW}}}(d>t_0) = 36.0$~nm, while $t_{{\rm{DW}}}(d \le t_0) = d$, $t_0=36$~nm. Curves 3 and 4 show the resistance of the PC with and without DW for the combined case, respectively. Other parameters are the same as in Fig.\ref{fig2}. The inset shows a zoom of the resistance oscillation of the single DW in linear scale.}
\label{fig3}
\end{figure*}
\begin{figure*}[!t]
\centering
\includegraphics[clip=true, width=4.8in]{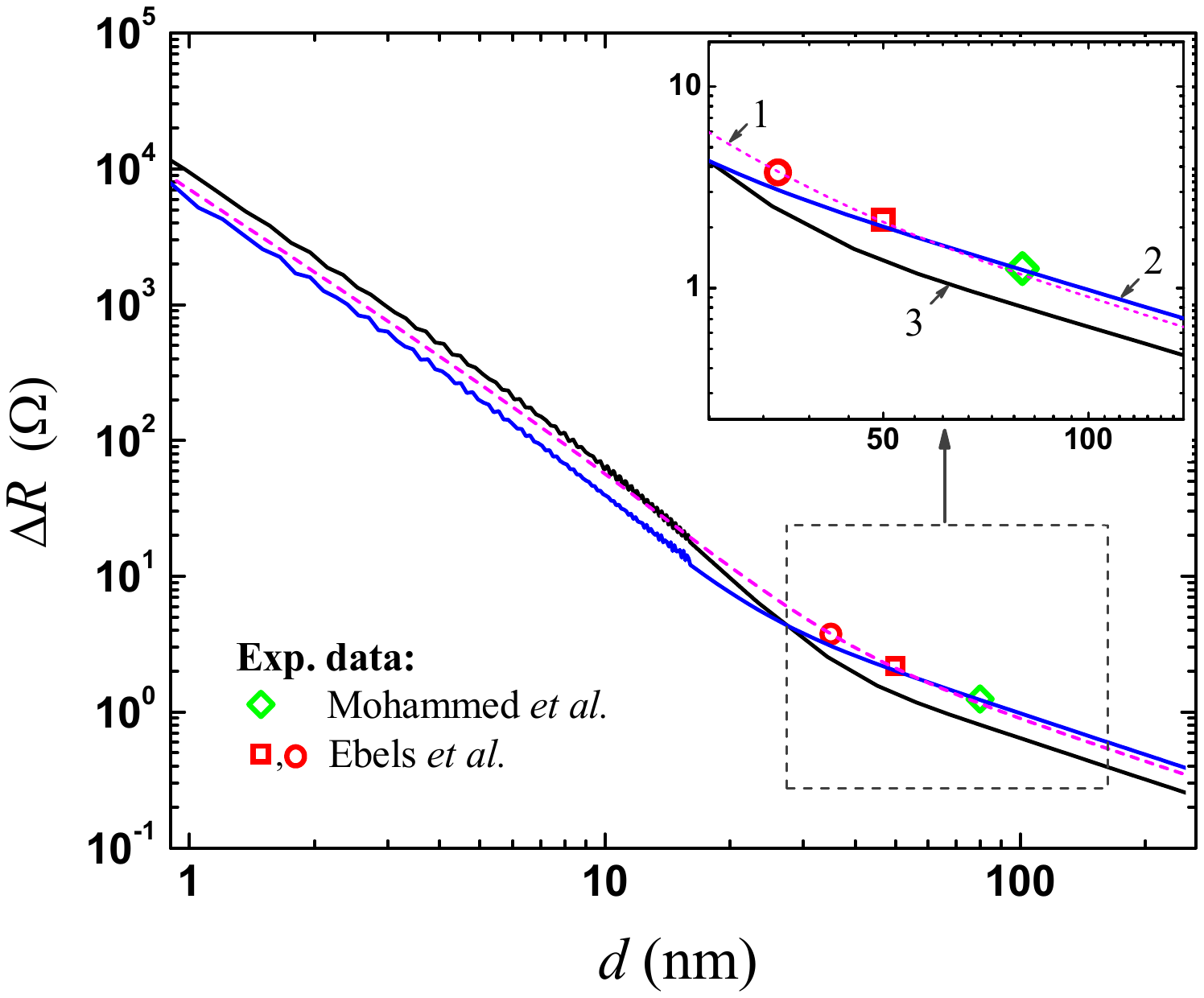}
\caption{Single DW resistance {\it{vs.}} diameter of the asymmetric PC, corresponding to the DW resistance in segmented NWs at $d \to d_0$. Parameters for the curve 2 is oriented for Co/Ni interface: $t_0=16$~nm, $k_ \uparrow ^L  = 0.61$~\AA$^{-1}$, $k_ \downarrow ^L  = 1.08$ ~\AA$^{-1}$, $k_ \uparrow ^R  = 0.65$~\AA$^{-1}$, $k_ \downarrow ^R  = 1.0801$~\AA$^{-1}$, $\ell _\uparrow^L= 12.0$~nm, $\ell_\downarrow ^L= 4.0$~nm, $\ell_\uparrow ^R = 6.0$~nm, $\ell _ \downarrow ^R= 1.5$~nm, having $\ell _ \uparrow ^L /\ell _ \downarrow ^L  = 3.0$ and $\ell _ \uparrow ^R /\ell _ \downarrow ^R  = 4.0$; Parameters for the curve 3: $t_0$, $k_ \downarrow ^L$, $k_ \uparrow ^R $, $k_ \downarrow ^R$, $\ell _ \uparrow ^L$ and $\ell _ \downarrow ^L$ are the same as for the curve~2,  while $k_ \uparrow ^L  = 0.41$~\AA$^{-1}$, $\ell _ \uparrow ^R= 5.8$~nm, $\ell _ \downarrow ^R  = 1.95$~nm and $\ell _ \uparrow ^{L(R)} /\ell _ \downarrow ^{L(R)}  = 3.0$. Dashed curve 1 is symmetric case of the Co-based PC, that coincides with curve 1 in Fig.\ref{fig3}.}
\label{fig4}
\end{figure*}
In contrast, segmented NWs with periodical Co/Ni interfaces naturally have initially different $k_s^{L}$ and $k_s^{R}$ values. As a result of fitting with an experimental data in the section below, the following initial $k_s$ and $l_s$ are found for Co NW: $k_ \uparrow ^L  = 0.61\,{\rm{\AA}}^{-1}$, $k_ \uparrow ^R  = 0.6101\,{\rm{\AA}}^{ - 1}$, $k_ \downarrow ^L  = 1.08\,{\rm{\AA}}^{-1}$, $k_ \downarrow ^R=1.0801\,{\rm{\AA}}^{-1}$, and $\ell_\uparrow ^{L(R)} = 12.0$~nm, $\ell_ \downarrow ^{L(R)}  = 3.0$~nm, however, $\ell_ \downarrow ^{L}$ was changed to $\ell_ \downarrow ^{L}=4.0$~nm for a better matching for the segmented Co/Ni NW, that, probably, related with a different lattice properties in relation to uniform Co NW.  The conduction band spin polarization parameter $\delta _{{\rm{Co}}}  = k_ \uparrow  /k_ \downarrow$ is equal to 0.56, that numerically coincides with the one in Ref. \cite{tagirov_quasiclassical_2007}, noting that simulations in present work give the same results making the complete spin index reversal $\downarrow(\uparrow)$ to $\uparrow (\downarrow)$ for each material parameter. The additional sources of experimental estimations of the spin-resolved Fermi wavenumbers and theoretical simulations of the averaged $\ell$ in different metals are available in Ref.\cite{himpsel_electronic_1999} and Ref.\cite{gall_electron_2016}, respectively.

\section{Results and Discussions}

The result of calculations for the DW resistance behaviors are shown in Fig.\ref{fig2}, $t_0=10$~nm. There are two cases for the fixed ($t_{{\rm{DW}}}(d) = t_0$ at any $d$, see curve 1) and combined DW width dependencies ($t_{{\rm{DW}}}(d) = d$ at $d \le t_0$ while $t_{{\rm{DW}}}(d) = t_0$ at $d > t_0$ for curve 2), respectively. Theoretical curves, considered for comparison with each other and experimental data, are shown in Fig.\ref{fig2}a and Fig.\ref{fig2}b. Experimental points are obtained as $\Delta R = 0.3\,\Omega$ at $d_0=300$~nm for Permalloy (Py) Ni$_{80}$Fe$_{20}$ NWs according Wong {\it{et al.}}\cite{wong_current-induced_2016}; $\Delta R = 1.2\,\Omega$ at $d_0=80$~nm by Mohammed {\it{et al.}} \cite{mohammed_angular_2015,mohammed2017} for Co/Ni segmented NW; $\Delta R = 2.178\,\Omega$ at $d_0=50$~nm, and finally, $\Delta R = 7.5\,\Omega$ at $d_0=35$~nm for uniform Co NWs by Ebels {\it{et al.}}\cite{ebels_spin_2000}. The last point for $d_0=35$~nm is considered as a result of two DW’s contributions, since a single DW resistance at this dimension is expected to be a half of $7.5\,\Omega$, {\it{i.e.}} $\Delta R = 3.75\,\Omega$, noting that Ebels {\it{et al.}} also assumed the presence of the doubled DW at this point. In case of a few DWs, existing in NW, the related DW resistance is a few times higher, respectively. Two points A and B in Fig.\ref{fig2}d demonstrate the relative 10\% deviation of the curves 1 and 2 from each other due to the oscillation of the curve 2. These deviations are suppressed at $t_{{\rm{DW}}}(d) > 10$~nm and raised due to presence of the DW at the confined geometry of PC, {\it{e.g.}} Fig.\ref{fig2}c clearly shows the related  $R_{{\rm{DW}}}$ oscillations with $d$.

The calculated $\Delta \bar R(d)$ and $\Delta R(d)$ - curves 1 and 2 for the fixed $t_{\rm{DW}}$ and combined cases are also shown in Fig.\ref{fig3}, their minor difference due to logarithmic scale is more clearly highlighted in the inset with liner scale. The black curve 2 is calculated in such a way to consider possible large value of $t_0 = 36.0$~nm to see the condition, where oscillations will be over. Figure \ref{fig3} includes the curves for PC’s resistances with presence of the DW $R_{{\rm{DW}}}$ - blue curve~3, and without $R_0$ - orange dash-dotted curve 4, correlating with the given behavior of the curve~2.
It is obvious, that the reason of the $\Delta R$ oscillations is $R_{{\rm{DW}}}$ variations. Its origin is the MFP dimensional effects due to ballistic and quasi-ballistic conditions of the electron scattering, where related barrier transparency {\it{vs.}}  $t_{{\rm{DW}}}$ behave dissimilar in relation to clear ballistic, or diffusive conditions. The most obvious range of the step-like oscillations is $\Delta R \approx 10 \,\Omega - 300\,\Omega$ at around $d \approx 2.5$~nm$-16.0$~nm, that is comparable with the range between minimal $\ell_{\rm{min}}=\ell_ \downarrow$ and maximal $\ell_{\rm{max}}=\ell_ \uparrow$ values. In particular, for the case $l_{\rm{min}}<t_{\rm{DW}}<l_{\rm{max}}$, the electron scattering on the DW is intermixed by two conductive spin channels at two different conditions: ballistic for electrons with spin up, while quasi-ballistic condition takes place for electrons spin down. The amplitude of oscillation is proportional to the difference between $l_{\rm{max}}$ and $l_{\rm{min}}$.
It should be noticed that observed $\Delta R$ steps can be used in fabrication for the minimization of the NW size distribution influence on a dispersion of the total resistance in NWs arrays, {\it{e.g.}} the convenient range of the stable $\Delta R$ value is $d \approx 10.4$~nm$-11.2$~nm, where $\Delta R = 45.1\,\Omega$ which the third step from the left side of the inset, Fig.\ref{fig3}. The oscillation averaged period is almost a constant value, that is around 1.73~nm at $d = 2.1$~nm$-15.2$~nm, while the range $d = 15.2$~nm$-36.0$~nm ($d \approx 1.5\ell _{\rm{max}}  - 3\ell _{\rm{max}}$) is different since the transport is more diffusive: the period of the oscillations monotonically decreases with a suppressed amplitude here. An important sensitive parameter for the system is the MFP ratio $\ell _ \uparrow ^{L(R)} /\ell _ \downarrow ^{L(R)}$. For the curves in Fig.\ref{fig3}, it is fixed for simplicity as $\ell _ \uparrow ^{L(R)} /\ell _ \downarrow ^{L(R)}  = 4.0$, being oriented mainly for homogeneous Co NW, rather than for the Py and Co/Ni cases. The estimation of the giant ${\rm{BMR}}$, which can be determined as ${\rm{BMR}} = \left( {R_{{\rm{DW}}}  - R_0 } \right)/R_0  \cdot 100\,\%$, gives $124\,\%$ for $d=1.2$~nm. The point at $d=350$~nm for Py NW is nicely fitted by $\ell _ \uparrow ^{L(R)} /\ell _ \downarrow ^{L(R)}  = 4.5$ (not shown), while the point $d = 80$~nm for Co/Ni NW is more precisely fitted by segmented NW's model, described below. 

In addition to the homogeneous case, it is worth to consider segmented NW (or asymmetric PC), where spin-resolved $k_s$ and $\ell_s$ values are different for the left and right sides of the interface between segments. It was found that the oscillation of $\Delta R$ also take place in this case for a constrained DW, however, its amplitude are much less in relation to homogeneous one, and its period is not monotonic, Fig.\ref{fig4}. The asymmetry is divided on two cases: the first when the curve 2 shows  $\Delta R$ for the both asymmetries by $k_ \uparrow  /k_ \downarrow$ and $\ell _ \uparrow  /\ell _ \downarrow $ ratios in the left and right sides, and second case when the curve 3 corresponds to the asymmetry by the $k_ \uparrow  /k_ \downarrow$ ratio only. In contrast to curve~2, curve~3 has larger difference between $k_ \uparrow ^L $ and $k_ \downarrow ^L$, its $\Delta R(d)$ is higher in the ballistic range and lower in the diffusive one. Moreover, there are diameters at the ballistic range for the curve~3, at which $\Delta R$ oscillations are suppressed, that is not well observed for the curve~2. The blue curve~2 in the diffusive regime with $\ell _ \uparrow ^L /\ell _ \downarrow ^L  = 3.0$ and $\ell _ \uparrow ^R /\ell _ \downarrow ^R  = 4.0$ better fits experimental point for the Co/Ni segmented NW at $d_0=80$~nm (green rhombus) than the case of homogeneous Co NW $\ell _ \uparrow ^{L(R)} /\ell _ \downarrow ^{L(R)}  = 4.0$  (curve 1), see the inset in Fig.\ref{fig4}. As a result, curve~2 and curve~3, representing asymmetric cases, additionally proof that $\ell_s$ ratios are the most valuable parameters in $\Delta R(d)$ behavior. This property can be used, for example, in simulation and fabrication of the magnetic interconnects, DW memory, {\it{etc}}, decreasing unwanted resistance dispersion. Raw data and program code are available online \cite{rep27}.

\section{Conclusion}
The adapted point-like contact model allows successfully describe experimental data of the single and double DW resistances in magnetic junctions and cylindrical NWs, when diameters of the NW and PC are close values. Almost four orders $\Delta R$ drop is held within only two orders by $d$, {\it{e.g}} theory predicts 7 k$\Omega$ for $d\approx 1$~nm and 0.88 $\Omega$ for $d\approx 100$~nm. The system is sensitive to the ratios of the spin-resolved electron MFP and wavenumbers of the magnetic material. The reason of the found $\Delta R$ oscillations with $d$ is the flexible DW approach when $d=t_{\rm{DW}}$ and take place the case of $l_{\rm{min}}<t_{\rm{DW}}<l_{\rm{max}}$ at which electron scattering on the DW has intermixing conditions: ballistic for one electron spin orientation and quasi-ballistic for another. The presented $\Delta R$ oscillations for Co nanowire, where the relative amplitude of oscillations does not exceed 10\%, are not related to quantization phenomena, or tilting of the DW-induced potential at  $t_{\rm{DW}}>t_{0}$, amplitude of oscillations vanishes when the system approaches $t_{\rm{DW}}>1.25\; l_{\rm{max}}$ and competition between ballistic and quasi-ballistic regimes is over. The $\Delta R$ steps can be used for the minimization of the total resistance dispersion coming from the size distribution of a large amount of NWs, it may help to detect approximate NW's dimension measuring only its resistance with and without DW, and support a further development of the racetrack memory concept.

\section*{Data Availability}
The raw data and program code: \url{https://data.mendeley.com/datasets/kmsjt7kndk/1}.

\section*{Acknowledgment}
This work was financially supported by the "Center for the Semiconductor Technology Research" from The Featured Areas Research Center Program within the framework of the Higher Education Sprout Project by the Ministry of Education (MOE) in Taiwan. Also supported in part by the Ministry of Science and Technology, Taiwan, under Grant MOST 110-2634-F-009-027-  and MOST 110-2112-M-A49-016- 

\end{multicols}
\end{document}